\pdfoutput=1

\documentclass[11pt]{article}

\usepackage[final]{acl}
\usepackage{times}
\usepackage{latexsym}
\usepackage{adjustbox}
\usepackage[T1]{fontenc}
\usepackage{listings}

\usepackage[utf8]{inputenc}
\usepackage{booktabs}
\usepackage{microtype}

\usepackage{inconsolata}
\usepackage{tabularray}

\usepackage{graphicx}
\usepackage{amsmath}
\usepackage{array}

\title{VoxRAG: A Step Toward Transcription-Free RAG Systems in Spoken Question Answering}

\author{
 \textbf{Zackary Rackauckas\textsuperscript{1}},
 \textbf{Julia Hirschberg\textsuperscript{1}},
\\
 \textsuperscript{1}Columbia University,
\\
 \small{
   \textbf{zcr2105@columbia.edu, julia@cs.columbia.edu}
 }
}

\begin{document}
\maketitle
\begin{abstract}
    We introduce VoxRAG, a modular speech-to-speech retrieval-augmented generation system that bypasses transcription to retrieve semantically relevant audio segments directly from spoken queries. VoxRAG employs silence-aware segmentation, speaker diarization, CLAP audio embeddings, and FAISS retrieval using L2-normalized cosine similarity. We construct a 50-query test set recorded as spoken input by a native English speaker. Retrieval quality was evaluated using \textit{LLM-as-a-judge} annotations. For very relevant segments, cosine similarity achieved a Recall@10 of 0.34. For somewhat relevant segments, Recall@10 rose to 0.60 and nDCG@10 to 0.27, highlighting strong topical alignment. Answer quality was judged on a 0–2 scale across relevance, accuracy, completeness, and precision, with mean scores of 0.84, 0.58, 0.56, and 0.46 respectively. While precision and retrieval quality remain key limitations, VoxRAG shows that transcription-free speech-to-speech retrieval is feasible in RAG systems.
\end{abstract}

\section{Introduction}

Traditional question-answering (QA) retrieval-augmented generation (RAG) systems retrieve text documents from a vector database by performing semantic similarity search from a user’s query. A large language model (LLM) then generates context-aware answers based on the retrieved content \cite{Rackauckas_2024}. This architecture, however, can be extended to operate directly on spoken audio instead of text. Retrieving spoken audio documents without relying on intermediate transcriptions is an emerging area of RAG research \cite{min2025speechretrievalaugmentedgenerationautomatic}.

We present VoxRAG: a modular, open-source retrieval pipeline for RAG with full speech-to-speech retrieval. Unlike hybrid text and audio systems, VoxRAG keeps both the user query and retrievable documents in audio form up to the generation stage, using Contrastive Language-Audio Pretraining (CLAP) embeddings \cite{elizalde2022claplearningaudioconcepts} to retrieve semantically relevant segments directly from podcast audio (see Appendix \ref{sec:qapairs} for sample QA pairs).

Using podcasts as a retrieval target presents challenges such as informal language, overlapping speakers, non-speech audio (e.g., music, laughter), and generally poor automatic speech recognition (ASR) transcription output quality \cite{jones2021currentchallengesfuturedirections}. VoxRAG mitigates these issues with silence-aware segmentation, speaker diarization, and CLAP embedding retrieval, avoiding early commitment to potentially faulty transcripts and enabling semantically grounded retrieval in the acoustic domain. We evaluate both retrieval and answer quality using RAGElo's \textit{LLM-as-a-judge} methods, which have shown positive alignment with human judgments in QA evaluation, to assess how well retrieved audio supports answer generation \cite{rackauckas2024evaluatingragfusionrageloautomated}.

Related work has explored RAG systems for audio in both text and hybrid modalities. The TREC 2020–21 Podcasts Track saw systems using ASR text retrieval and summarization \cite{clifton-etal-2020-100000}, including fine-tuned BART \cite{lewis2019bartdenoisingsequencetosequencepretraining} and Whisper spoken term detection. Hybrid systems like Schwertzer’s combination of COLA \cite{saeed2020contrastivelearninggeneralpurposeaudio} and RoBERTa \cite{liu2019robertarobustlyoptimizedbert} show promise in mixed-modal retrieval \cite{schwerter2022podcast}.

More recent models embed audio and text into shared or comparable vector spaces. SpeechDPR distills from ASR and dense passage retrieval (DPR) systems to embed spoken passages directly \cite{lin2024speechdprendtoendspokenpassage}, while SEAL uses separate encoders for speech and text to enable cross-modal retrieval without transcription \cite{sun2025sealspeechembeddingalignment}. Spectron processes spectrograms for QA entirely within an LLM framework \cite{nachmani2024spokenquestionansweringspeech}, and SpeechRAG integrates speech retrieval with an LLM for answering text queries from raw audio \cite{min2025speechretrievalaugmentedgenerationautomatic}. Meanwhile, DUAL demonstrates fully speech-native retrieval by embedding discrete speech units without paired text training \cite{lin2022dualdiscretespokenunit}. VoxRAG contributes to this emerging space by exploring a retrieval-first, speech-native architecture that maintains audio representations up to the point of answer generation, differing from span-prediction models like DUAL.

\section{Method}

Our construction of VoxRAG was motivated by two core research questions: 1) Can we retrieve semantically relevant documents directly from spoken language and without relying on text representations? 2) Can those documents support high-quality answer generation using an LLM?

We define “high-quality” segments as those that contain very relevant or somewhat relevant information to a user's query. We define "high-quality" answers along four axes: relevance, accuracy, completeness, and precision.

\subsection{Podcast Indexing}

Each podcast is processed through a modular indexing pipeline with speaker diarization, silence-aware segmentation, audio embedding, and optional transcription. Diarization is handled via NeMo’s ClusteringDiarizer \cite{kuchaiev2019nemotoolkitbuildingai}, while speech segmentation uses Silero VAD \cite{silero_vad_2024}. Transcripts are generated using Faster-Whisper \cite{radford2022robustspeechrecognitionlargescale, faster_whisper} and are only used for LLM input and display rather than retrieval.

All speech segments are embedded using CLAP \cite{elizalde2022claplearningaudioconcepts}, which maps audio to a joint audio-language embedding space (see Appendix \ref{sec:hyperparameters}). This allows semantic-level retrieval even in the absence of exact word overlap, making it more robust for podcast audio that includes informal speech, background noise, or laughter. While traditional models like \textit{wav2vec 2.0} \cite{baevski2020wav2vec20frameworkselfsupervised} focus on phonetic or acoustic information, CLAP learns to associate audio with language in a shared space. This lets us treat podcast segments like paragraphs of meaning rather than waveforms or phonemes \cite{elizalde2022claplearningaudioconcepts} for direct audio retrieval.

\noindent
   \textbf{Audio Loading and Preprocessing: }
       Each podcast file is loaded, converted to mono, and resampled to 16 kHz. \\
   \textbf{Segmentation and Diarization: }
       Diarization is used to detect speaker turns and assign segment-level speaker IDs.
       VAD identifies valid speech spans, which are then merged with speaker labels to define segments. \\
   \textbf{Embedding and Optional Transcription: }
       Segments are embedded with CLAP and stored in memory.
       Transcripts are generated and aligned with segments for LLM prompting.

\subsection{Retrieval}

At query time, we take a spoken user query, process it through the same pre-processing and CLAP embedding pipeline, and compute cosine similarity between the query and all indexed segment embeddings using FAISS \cite{douze2025faisslibrary} (see Appendix \ref{sec:hyperparameters}). The top ten segments are selected as candidates. We evaluate two retrieval configurations: (i) cosine similarity only and (ii) cosine followed by the \textit{ms-marco-MiniLM-L6-v2} cross-encoder reranker. All other hyper-parameters are kept identical. Our primary analysis focuses on retrieval using cosine similarity, as shown in Table~\ref{tab:mrr5_scores}.

\noindent
   \textbf{Query Processing:}
       The user’s spoken query is loaded, normalized, and embedded using CLAP. \\
   \textbf{Similarity Search:}
       The top ten segments are retrieved by cosine similarity.
       Neighboring segments (before and after) are included for context.

\subsection{Answer Generation}

VoxRAG's modularity supports evaluation of chunking, embedding, and retrieval strategies. Once segments are retrieved, their transcripts are passed along with the transcribed query to \textit{GPT-4o} to generate a natural language response.

\noindent
   \textbf{Prompt Construction:}
       Retrieved segment transcripts are labeled with the speaker and the segment number.
       The transcribed query and these segments are formatted as a prompt for the LLM. \\
   \textbf{Generation and Display:}
       \textit{GPT-4o} returns a natural language answer.
       The answer is shown in a Gradio interface alongside audio players for each segment.
       
\section{Experiments}

\subsection{Dataset and Evaluation Queries}

We selected twenty episodes from the Trash Taste podcast as our source corpus. These episodes feature three main speakers: Joey, Connor, and Garnt, with occasional guest speakers. For our main evaluation, we used a single representative episode with a run time of 2 hours and 3 minutes. This episode was segmented into 202 chunks using silence-aware merging and speaker diarization, ensuring that each segment remained under 90 seconds in length. Although our evaluation focuses on a single episode, the system is capable of processing extended podcast archives comprising many hours of audio.

To evaluate the system's ability to handle real-world questions, we curated 11 organic queries from a Tokyo Weekender article titled "11 Questions With Anime Podcast Trash Taste"\footnote{https://www.tokyoweekender.com/tw-community/trash-taste-podcast/} and a live Trash Taste QA session \footnote{https://youtu.be/tzFLreIzB78?si=yI96MWYgvQdmspsl}. To expand the test set, we generated 205 synthetic queries using \textit{GPT-4o} and 294 using \textit{GPT-o1}. From these, we randomly sampled 50 non-duplicative synthetic queries for a final test set of 50 diverse, high-variance questions.

All text queries were then read aloud by the same male native English speaker in a controlled environment and recorded using Audacity.

\subsection{Retrieval Quality}

We evaluated retrieval performance with Recall@10 and normalized Discounted Cumulative Gain at 10 (nDCG@10). Following the RAGElo evaluation toolkit \cite{rackauckas2024evaluatingragfusionrageloautomated} (see Appendix \ref{sec:prompts}), we conducted two separate evaluations using \textit{LLM-as-a-judge} annotations, one where segments were labeled as either very relevant (1) or not relevant (0), and another where segments were labeled as somewhat relevant (1) or not relevant (0). This allowed us to assess precise retrieval performance and broader topical alignment.


\begin{table}[ht]
\centering
 \caption{Retrieval performance of VoxRAG using cosine similarity (with and without cross-encoder--CE--reranking) on very relevant (VR) and somewhat relevant (SR) documents.}
 \label{tab:mrr5_scores}
 \small
  \begin{tblr}{
    width=\columnwidth,
    colspec={X[l, c] X[l, c] X[l,c] X[l,c]},
    colsep=1pt,
    rowsep=0pt,
    row{odd}={bg=azure9},
    column{1} = {cmd=\small},
    row{1} = {c, bg=azure3, fg=white, font=\sffamily, cmd=\textbf},
    hline{1,Z} = {solid, 1pt},
  }
    Setup & Recall@10 & nDCG@10 \\
    Cosine (VR)         & 0.34 & 0.03 \\
    Cosine (SR)         & \textbf{0.60} & \textbf{0.27} \\
    Cos + CE (VR)       & 0.26 & 0.03 \\
    Cos + CE (SR)       & 0.46 & 0.14 \\
  \end{tblr}%
\end{table}

As shown in Table~\ref{tab:mrr5_scores}, cosine similarity with and without a reranker retrieves segments with modest absolute scores, consistent with the challenges of speech-to-speech retrieval. Based on the large improvement in somewhat relevant over very relevant documents, while the system often retrieves topically aligned audio, it struggles to consistently retrieve precise, direct answers. The large gap between very relevant and somewhat relevant retrieval scores underscores the difficulty of fine-grained semantic matching using current audio embeddings.

\subsection{Answer Quality}

We evaluated answer quality along four dimensions: relevance, accuracy, completeness, and precision, each rated on a 0–2 linear scale by an impartial LLM judge using \textit{GPT-4o}. This process follows the method of RAGElo \cite{rackauckas2024evaluatingragfusionrageloautomated} (see Appendix \ref{sec:prompts}). Each generated answer was based on the transcripts of the top ten retrieved audio segments, along with adjacent context for continuity. Table~\ref{tab:answer_quality} presents the mean scores and standard deviations for each evaluation dimension.

\begin{table*}[ht]
\centering
\small
 \caption{Mean answer quality scores (0--2 scale) from impartial LLM judges across evaluation dimensions. Relevance significantly outperforms all other metrics. Effect sizes ($d$) and $p$-values are computed relative to relevance using two-tailed paired $t$-tests.}\label{tab:answer_quality}
  \begin{tblr}{
    width=\textwidth,
    colspec={X[l,c] X[c] X[c] X[c] X[c] X[c] X[c]},
    colsep=2pt,
    rowsep=0pt,
    row{odd}={bg=azure9},
    column{1} = {cmd=\small},
    row{1} = {c, bg=azure3, fg=white, font=\sffamily, cmd=\textbf},
    hline{1,Z} = {solid, 1pt},
  }
  Metric & Mean & Std Dev & $\Delta$ vs. Relevance & $d$ & $p$-value & Significantly Lower \\
  Relevance     & 0.84 & 0.87 & ---   & ---  & ---     & ---   \\
  Accuracy      & 0.58 & 0.81 & -0.26 & 0.49 & $<$ 0.01  & Yes   \\
  Completeness  & 0.56 & 0.81 & -0.28 & 0.52 & $<$ 0.01  & Yes   \\
  Precision     & 0.46 & 0.81 & -0.38 & 0.67 & $<$ 0.01 & Yes   \\
  \end{tblr}%
\end{table*}

Relevance had the highest mean score (0.84), significantly outperforming all other dimensions ($p < 0.01$), with a medium-to-large effect size (Cohen’s $d = 0.67$) when compared to precision. This suggests that, while the system frequently retrieved content that was topically appropriate, it often failed to deliver factual specificity or grounding. In other words, the answers were often “about the right thing,” but lacked detail.

Completeness and accuracy were closely aligned, with means of 0.56 and 0.58 respectively ($p = 0.32$, $d = 0.14$), implying that partially correct answers were also seen as incomplete. Precision received the lowest average score (0.46) and was significantly lower than all other metrics. The model often failed to refer to the correct episode, moment, or speaker with sufficient granularity.

A correlation analysis reinforced these findings. Accuracy, completeness, and precision were all tightly linked ($r > 0.91$), suggesting that they capture a shared dimension of factual correctness and detail. Relevance, by contrast, was more loosely correlated with the others ($r \approx 0.77$), supporting the idea that being on-topic alone is insufficient for generating high-quality responses.

Despite the general trend, ten queries, or 20\% of all queries, achieved perfect scores across all dimensions, suggesting that when embedding alignment and segment selection succeed, VoxRAG delivers strong results. A comparatively large number of queries containing the word “shower” received perfect scores, though this was not consistent across all such queries. Of the ten queries containing “shower,” four achieved perfect marks, as compared to 20\% overall. While anecdotal, this partial pattern may still reflect idiosyncrasies in how CLAP embeddings handle certain personal or lifestyle-related concepts potentially influenced by consistent acoustic or contextual cues in the training data.

\section{Discussion}

Our results highlight the challenges inherent in speech-to-speech retrieval within unstructured, multi-speaker podcast content. Although CLAP embeddings provided a degree of coarse semantic alignment, the retrieval process often prioritized topically related segments over precise matches. This tendency resulted in lower precision and incomplete responses. Evaluations revealed strong correlations between accuracy, completeness, and precision, indicating a shared reliance on fine-grained factual grounding. In contrast, relevance scores remained consistently high, suggesting that topical alignment alone is insufficient for generating high-quality answers. Certain queries, particularly those involving lifestyle concepts such as "shower," achieved perfect scores in some cases, but not reliably. This inconsistency may reflect variability in how well specific topics are represented within audio embeddings and warrants further investigation. 

The modular architecture of VoxRAG enabled rapid experimentation across embedding models, chunking strategies, and retrieval logic. The inclusion of audio playback within the interface proved valuable for error analysis, as it revealed retrieval mismatches that were not apparent from text alone. 

These findings establish a baseline for future research on audio-native question answering. They point to the need for improved embedding fine-tuning, more effective segmentation methods, and reranking strategies that better reflect factual precision. VoxRAG represents a step toward multimodal RAG systems capable of operating directly on real-world, noisy, and informal spoken content. With the proliferation of audio media, systems of this kind will be increasingly important for enabling direct retrieval and reasoning over speech, without dependence on textual transcripts.

\section{Conclusion}

VoxRAG explores the viability of a fully speech-to-speech retrieval pipeline for retrieval-augmented generation. While the system ultimately produces text answers, it retrieves documents directly from audio using CLAP embeddings \cite{elizalde2022claplearningaudioconcepts}, bypassing early transcription. Despite its novel architecture, the system underperforms on precision, completeness, and accuracy metrics, highlighting the limitations of current audio embedding models for fine-grained semantic retrieval. However, the retrieval quality on certain queries demonstrates the ultimate viability of RAG with speech-to-speech retrieval.

\section*{Limitations}

While VoxRAG shows that transcription-free audio-to-audio retrieval is feasible, several challenges remain. One key limitation is the absence of transcript-based or hybrid retrieval baselines. We do not compare against methods like CLAP with transcribed input or strong text retrievers such as BM25, which makes it difficult to assess the true tradeoffs of avoiding transcription. Another issue lies in the hybrid nature of the pipeline. While retrieval is audio-only, the system still relies on Whisper transcripts for answer generation, reintroducing ASR noise and undercutting the goal of being fully transcription-free. Future work should explore audio-native generation methods.

We also note potential bias in evaluation. GPT-4o is used for both generating and assessing answers, which may lead to overestimation of performance due to model self-agreement. Using a different model, such as Qwen or Mistral, for evaluation could help mitigate this. Our evaluation is further limited by the use of only one episode from the Trash Taste podcast, restricting diversity and generalizability. Broader testing across multiple episodes and speakers would provide stronger insights.

Finally, the system shows a gap between topical relevance and factual precision. CLAP embeddings retrieve on-topic segments, but these often lack the detailed grounding needed for accurate answers. Improving fine-grained alignment remains an open challenge. These limitations are expected at this early stage and help clarify where future work can focus to strengthen audio-native retrieval and QA pipelines.

\bibliography{custom}

\appendix

\section{Evaluator Prompts}
\label{sec:prompts}

\subsection{Retrieval Evaluator}
We used the following system prompt for our retrieval evaluator for very relevant documents:

\lstset{
  basicstyle=\ttfamily\small,
  breaklines=true,
  frame=none,
  numbers=none,
  showstringspaces=false
}

\begin{lstlisting}
You are an expert annotator evaluating whether a *spoken podcast transcript segment* is *very relevant* to a user's question.
These transcripts may include humor, casual speech, tangents, or non-traditional structure.
Return **1** if the segment contains strong, clear, and direct information addressing the user's question. 
Return **0** if the segment is only loosely or partially related, or entirely off-topic. 
Only respond with a single digit: 1 or 0. Do not explain.
\end{lstlisting}

For evaluating somewhat relevant documents, we used the following prompt:

\begin{lstlisting}
You are an expert annotator evaluating whether a *spoken podcast transcript segment* is *somewhat relevant* to a user's question.
These transcripts may include humor, casual speech, tangents, or non-traditional structure.
Return **1** if the segment has a loose or minor connection to the user's question - it may touch on a related theme, mention something adjacent, or vaguely resemble the topic, even if it is incomplete or off-target.
Return **0** if the segment has no real connection at all.
Only respond with a single digit: 1 or 0. Do not explain.
\end{lstlisting}

\subsection{Answer Evaluators}
For the answer quality evaluation, we used the following prompt:

\begin{lstlisting}
You are an impartial judge for evaluating the quality of the responses provided by an AI assistant tasked with answering users' questions about the *Trash Taste* podcast.

You will be given the user's question and the answer produced by the assistant. The assistant's answer was generated based on a set of audio-derived documents retrieved from episodes of the *Trash Taste* podcast.

You will be provided with the relevant podcast segments retrieved by the search engine.

Your task is to evaluate the answer's quality based on the response's **relevance**, **accuracy**, **completeness**, and **precision**, grounded in the retrieved podcast content.

## Rules for evaluating an answer:
- **Relevance**: Does the answer address the user's question?
- **Accuracy**: Is the answer factually correct, based on the retrieved podcast segments?
- **Completeness**: Does the answer provide all the information needed to address the user's question?
- **Precision**: If the user asks about a specific episode, moment, guest, or topic, does the answer correctly identify and reflect that specific context?

## Steps to evaluate an answer:
1. **Understand the user's intent**: Restate what the user is trying to find out, in your own words.
2. **Check if the answer is correct**: Think step-by-step about whether the answer truthfully and fully responds to the user's question.
3. **Evaluate the quality of the answer**: Judge the answer on relevance, factual accuracy (according to the retrieved podcast segments), and how completely it covers the query.
4. **Assign a score**: Produce a single-line JSON object with the following keys, each with a score from 0 to 2:

- "relevance"  
  - 0: The answer is not relevant to the user's question.  
  - 1: The answer is partially relevant.  
  - 2: The answer is fully relevant.

- "accuracy"  
  - 0: The answer is factually incorrect or contradicts the retrieved content.  
  - 1: The answer is partially correct but includes errors or misinterpretations.  
  - 2: The answer is factually correct based on the retrieved segments.

- "completeness"  
  - 0: The answer leaves out major parts of the question.  
  - 1: The answer addresses the question only in part.  
  - 2: The answer covers all key aspects of the user's question.

- "precision"  
  - 0: The answer refers to the wrong episode, topic, or context.  
  - 1: The answer is somewhat related but not specific enough.  
  - 2: The answer directly reflects the specific content or moment asked about.

The last line of your evaluation must be a SINGLE LINE JSON object with the keys "relevance", "accuracy", "completeness", and "precision", each assigned a score between 0 and 2.

[ DOCUMENTS RETRIEVED ]  
{documents}

[ User Query ]  
{query}

[ Agent Answer ]  
{answer}
\end{lstlisting}

\section{Hardware and Runtime Configuration}
\label{sec:hardware}
Experiments were conducted on two machines: one with an RTX 5090 GPU, and another with two NVIDIA A4500 GPU's. On the latter machine, from user input to LLM output, the average generation time was 3.02 seconds. Under these conditions, the system demonstrates practical viability for offline indexing and interactive, low-latency query-time retrieval.

\section{Embedding Hyperparameters}
\label{sec:hyperparameters}
Audio segments and spoken queries were embedded using the \textit{laion/clap-htsat-unfused} model. Prior to indexing and similarity search, both query and document embeddings were L2-normalized. This normalization step ensures that inner product similarity in the FAISS \texttt{FlatIP} index is equivalent to cosine similarity.
This enables semantically aligned retrieval directly in the CLAP embedding space without requiring additional reranking or feature engineering.

\section{Sample QA Pairs}
\label{sec:qapairs}
\begin{table}[ht]
    \centering
    \caption{A sample of QA pairs on Synthetic Queries}\label{tab:base_queries}
    \begin{tblr}{
      width=\columnwidth,
      colspec={X[l,1]},
      colsep=1pt,
      rowsep=0pt,
      row{odd}={bg=azure9},
      column{1} = {cmd=\small},
      row{1} = {c, bg=azure3, fg=white, font=\sffamily, cmd=\textbf},
      hline{1,Z} = {solid, 1pt},
    }
    QA pairs (Q = white, A = blue)\\
    \textbf{Q:} Can the one-minute shower routine ever replace your normal shower, or is that impossible?\\
    \textbf{A:} ... Connor expresses skepticism about whether such a quick shower can truly replace a normal shower. He mentions that taking a shower is not just about getting clean but also about the time it takes to refresh and wake up properly. Joey adds that a shower under five minutes might not be sufficient to adequately clean oneself, suggesting that such a quick routine is more about covering up body odor than truly cleaning. Essentially, both seem to agree that while a one-minute shower might serve in a pinch, it cannot fully replace the benefits of a longer, more thorough shower. This perspective is captured in Segment 17. \\
    \textbf{Q:} What travel conditions now make or break a trip for you? \\
    \textbf{A:} ... Garnt shared that he finds Kyoto to be a very touristy spot, which can detract from the experience. He mentioned that while it's worth visiting at least once, the crowds and the commercialization of the area can take away from its magic. He expressed a preference for the outskirts of Kyoto, which tend to be more enjoyable and less crowded (Segment 20). Connor added that the disparity between Kyoto's historical significance and its touristy nature can make visits feel less authentic. He indicated that while it might be great for first-time visitors, the experience might not be as enjoyable upon return due to the overwhelming number of tourists (Segment 20). Overall, both speakers highlight that the level of crowding and commercialization in popular tourist destinations can make or break the quality of a travel experience. \\
    \textbf{Q:} If you made a Trash Taste Bingo Card, what squares would be absolutely required?\\
    \textbf{A:} ...
    1. Mention of League of Legends — This game often comes up in discussions (Segment 2).
    2. Reference to drinking stories — The guys frequently share hilarious and embarrassing drinking experiences (Segment 5).
    3. Discussion about sibling dynamics — They talk about their experiences growing up as siblings or only children (Segment 26).
    4. Anime recommendations — Each episode usually involves recommendations or debates about anime (Segment 19).
    5. Joey's love for food — Food discussions, especially around Japanese cuisine, are a staple (not specifically cited, but a recurring theme).
    6. Mention of gaming frustrations — Conversations about the challenges of gaming, especially in competitive environments (Segment 17).
    7. Cultural references — They often reference their cultural backgrounds, especially regarding Australia and the UK (Segment 10) ...
    \end{tblr}
\end{table}

\end{document}